\documentstyle[12pt,cite]{article}
              
\title{On the Observables Describing a Quantum Reference Frame.}
\author{S. Mazzucchi  \\ 
Department of Physics of the University \\
I-38050 Trento, Italy}

\newtheorem{proposition}{Proposition}
\newtheorem{definition}{Definition}
\begin{document} 
\maketitle                             
                 
\begin{abstract}
A reference frame $F$ is described by the element $g$ of the Poincar\'e group $\cal P$ which connects $F$ with a given fixed frame $F_0$. If $F$ is a quantum frame, defined by a physical object following the laws of quantum physics, the parameters of $g$ have to be considered as quantum observables. However, these observables are not compatible and some of them, namely the coordinates of the origin of $F$, cannot be represented by self-adjoint operators.  Both these difficulties can be overcome by considering a positive-operator-valued measure (POVM) on $\cal P$, covariant with respect to the left translations of the group, namely a covariance system. We develop a construction procedure for this kind of mathematical structure. The formalism is also used to discuss the quantum observables measured with respect to a quantum reference frame. 
\bigskip  \bigskip

\end{abstract}

\section{Introduction}

As Mach remarked at the end of the nineteenth century \cite{Mach}, from the physical point of view a frame of reference is defined by a material object of the same nature as the objects that form the system under investigation and the measuring instruments. Such an idea doesn't conflict with classical mechanics: for example a rigid body can define a spatial origin and an orientation. The situation becomes more complicated in quantum mechanics: Heisenberg's uncertanty relations forbid the exact determination of the position and the velocity of a frame. As noted by some authors \cite{AK}, such an analysis not only contributes to remove a classical concept from quantum mechanics, but also reveals some surprising physical consequences, such as the ``paradox of the quantum frames". That is, if we have three frames of reference $F_1,F_2$ and $F_3$, the observables which describe the relation between $F_1$ and $F_2$ may be not compatible with the observables which describe the relation between $F_2$ and $F_3$, even if the systems don't interact.

We follow an operational approach: the mathematical structures involved should have a direct physical meaning. From this point of view, a frame is determined by the procedures which transform an initial frame into the chosen one.
The set of the transformations allowed by a relativistic theory is represented by the proper orthochronous Poincar\'e group $\cal P$. Each element of $\cal P$ can be identified by means of ten indipendent parameters, indicating the coordinates of the new origin in the Minkowsky space-time, the three components of the velocity and three angles of orientation. From a physical point of view these ten variables can be determined by their  measurement performed on the physical object defining the frame: They have to be considered observables.
Unfortunately spectral measures, the mathematical structures traditionally associated to the physical concept of observables in quantum mechanics, cannot describe neither simultaneous measurements of position and velocity, nor measurements of time (Pauli's theorem). According to Gleason's theorem, the natural generalization of spectral measures, compatible with the ``Copenhagen interpretation'', is given by the so-called {\it positive operator value measures} (POVMs) \cite{{Busch},{Davies},{Holevo}}. \par
An observable is often characterized by its transformation properties under a particular symmetry group. We can define a {\it system of covariance} as a POVM endowed with its covariance properties under some symmetry group \cite{{Cattaneo},{CH},{Werner1}}. 
 If a POVM reduces to a spectral measure, the structure so defined is called a{\it system of imprimitivity} \cite{Mackey4}.\\
Following some hints which can be found in ref \cite{Werner1}, in section 2 we illustrate a general construction procedure for covariant observables, which allows one to assign the statistical distribution of the outcomes to the state vector of the system on which the measurement is performed. The cornerstones of our procedure are three theorems. The first one (covariant dilatation) asserts that any system of covariance can be derived from a suitable system of imprimitivity to which it is linked by means of a suitable ``intertwinig operator". The second theorem is Mackey's imprimitivity theorem, which allows us to find the most general form of a system of imprimitivity. The third one is the ``intertwinig operator theorem", which we derived in order to find the most general intertwining operator connecting the systems of imprimitivity to the unknown system of covariance. Its application is possible when the unitary representations of the symmetry group acting on the involved Hilbert spaces are decomposed into irreducible unitary representations.\par
We  stress that our results are very general and allow one to describe all the possible measurements of a given observable, defined by its spectrum and its trasformation properties under a relevant symmetry group, performed on a physical system which is identified by its covariance properties under the same group. We do not introduce any model, but use general symmetry properties of the measurement, to get all the POVMs describing a choosen observable.

In section 3, the developed procedure is used for a new derivation of the most general POVM on Minkowsky space-time which is covariant with respect to the Poincar\'e group, found in ref \cite{Toller1} by means of a different method.

Coming back to quantum frames in section 4, their description can be given by a system of covariance on the proper orthochronous Poincar\'e group, which is in this case both the parameter space and the symmetry group. Harmonic analysis on $SL(2C)$ and on the group of translations of ${\bf R}^4$ allows the decomposition of the most general unitary representation of $\cal P$ into irreducible unitary representations. In this way, the intertwining operator theorem can be applied and the most general probability distribution can be found. \\
A comparison with the Poincar\'e covariant POVM on Minkowsky space-time 
 indicates the existence of some constraints. In particular the so-called baricentric measures cannot be obtained: In other words one cannot require that the coordinates of the origin coincide with the coordinates of the centre of mass of the physical system defining the frame. Moreover through an analysis of our results, one realizes that, for a complete description of a quantum reference frame, kinematical variables are not sufficient and internal degree of freedom of the system have to be involved.\par
Finally, the formalism we adopted also allows an alternative derivation of the paradox of the quantum frames. Indeed, in section 5, we derive the form of the POVMs describing the relative observables between a generical quantum object and a quantum reference frame. The observables describing the relations between non-interacting quantum frames are just a particular case.\par
We  hope that an analysis of the variances of the probability distributions we have found will give a new class of indeterminacy relations. 
\section{Mathematical tools}
In the traditional framework of quantum mechanics, the states of a system are described by rays in a complex separable Hilbert space, or by normalized positive trace class operators, while observables are described by self-adjoint operators or, equivalentely, by spectral measures. It is well known that the last ones cannot describe neither joint measurements of incompatible observables, nor measurements of time, indeed Pauli's Theorem \cite{{Pauli},{Wightman}} forbids the description by means of a self-adjoint operator of an observable canonically conjugate to an Hamiltonian with a semibounded spectrum. Nevertheless, as noted by some authors \cite {{BGL2}, {BU1},{Werner1}}, the description of some measuring instruments requires a different mathematical structure, which can be recognized as a generalization of spectral measures: the so-called positive operator value measures (POVM). Gleason's theorem assures they are the most general mathematical structures describing observables compatible with the probabilistic interpretation of quantum mechanics. \\
The analysis of the proof of Pauli's theorem, shows that it is based on the covariance of time measurements with respect to time translations. This is not accidental, but shows the importance of simmetry in our discussion.
Indeed, the requirement of  precise covariance properties of the quantum  measurement under some simmetry group $\cal G$ leads to the following definition of covariance systems \cite{CH}.

\noindent{\bf Notation} From now on:
\begin{enumerate} \item $S$ will indicate a topological space which is locally compact and has a countable base of its topology. $S$ is called the ``space of the possible results of the measurement''.
\item $\cal H$ will indicate a complex separable Hilbert space. $\cal H$ is called ``the space of states (of the quantum system)''.
\item $\cal G$ will indicate a locally compact topological group which has a countable base of its topology. $\cal G$ is called the ``symmetry group of the theory''.
\end{enumerate}
\begin{definition} 
Let $\cal H$ be a space of states and $S$ be a space of possible results. A POVM on $S$ is a class $\tau := \{\tau (I)\}_{I\in {\cal B}(S)}$, 
where  $\cal B$ is the $\sigma $- algebra of Borel sets of $S$ and each $\tau (I) : {\cal H}\to{\cal H}$ is a positive bounded operator such that:
\begin{enumerate}
\item  $ \tau (I) \geq \tau (\emptyset  ) = 0 \qquad  \forall I \in {\cal B}$ 
\item  $\tau (\cup I_i)=\sum{\tau(I_i)}  $
\item  $\tau (S) =1$
\end{enumerate}
\noindent where $\{I_i\}$ is a countable collection of disjoint elements of ${\cal B}$ and the convergence is in the weak topology.
\end{definition}
Given a space of states $\cal H$, a space of possible results $S$ and a POVM $\{\tau (I)\}_{I\in {\cal B}(S)}$, for any pure state of the system determined by a normalized vector $\phi \in {\cal H}$, the probability that the outcome of the measurement of the observable described by $\{\tau (I)\}_{I\in {\cal B}(S)}$ belongs to the Borel set $I$ is:
\begin{equation}\label{probp}P(\phi , I)=\langle \phi, \tau (I)\phi \rangle.
\end{equation}
More generally, for any mixed state of the system determined by  a normalized positive trace class operator $\rho$ on $\cal H$, the probability above is given by:
\begin{equation}\label{probm}P(\rho , I)= {\rm Tr}[\rho \tau (I) ].
\end{equation}

\noindent{\bf Note: }
If $\tau (I_1\cap I_2)=\tau(I_1) \cdot \tau (I_2)$ for all $I_1,I_2\in {\cal B}$ then $\tau $ is a spectral measure.\par
The physical requirement that two observers, related by a transformation of the simmetry group $\cal G$ and performing the same experiment, get the same statistical distribution of the outcomes of the measurement, leads to a natural covariance condition and, eventually, to the following definition:
\begin{definition}
Let $\cal H$ be a space of states, $S$ be a space of possible results and $\cal G$ a symmetry group of the theory. Suppose that $\cal G$ acts on $S$ by means of a representation $\Lambda :g\to \Lambda (g),g\in {\cal G}$, where $\Lambda (g):S\to S$ are Borel mappings. Suppose $\cal G$ acts on $\cal H$ by means of a strongly continuous unitary representation $U:g\to U(g)$. Let $\{\tau (I)\}_{I\in{\cal B}(S)}$ a POVM on $S$, with the property:
\begin{equation}
U(g)\tau (I)U(g)^\dag=\tau(\Lambda (g)I)
\end{equation}
for any $I\in {\cal B}(S),g\in {\cal G}$. In this case the structure $({\cal H},S,{\cal G},\Lambda ,U,\tau)$ so defined is called system of covariance.
If, furthermore, $\tau$ is a spectral measure, it is called  system of imprimitivity \cite{{CH},{CM},{Cattaneo},{Mackey4},{Scutaru},{Werner1}}.
\end{definition}

While spectral measures represent a ``property" of the system on which the measurement is performed, generic POVMs can't describe ``definite observables'', but we have to prefer them because they are able to describe simultaneous measurements of incompatible observables and measurements of time.\par
While there is a unified treatment of imprimitivity systems, mainly due to G. W. Mackey, we cannot say the same for covariance systems. Anyway we can overcame this difficulty by means of the following theorem, which may be recognized as a covariant version of Naimark's dilatation theorem \cite{{Cattaneo},{Werner1}}. 
\begin{proposition} Let $({\cal H},S,{\cal G},\Lambda ,U,\tau)$ a system of covariance. Then there is an imprimitivity system $({\cal H'},S,{\cal G},\Lambda ,V,E)$, where $\cal H'$ is an Hilbert space, $V$ a strongly continuous unitary representation of the symmetry group $\cal G$ acting on $\cal H'$, $E$ is a spectral measure on the Borel $\sigma$-algebra $\cal B$ of $S$, and there is an ``intertwining operator" $A:{\cal H}\to {\cal H'}$, with the property $AU(g)=V(g)A,$ for any $g\in {\cal G}$, so that the following relation connects the spectral measure $E$ to the POVM $\tau$:
\begin{equation}
 \tau (I) = A^+E(I)A.
\end{equation}
Moreover, $\tau$ is normalized (i.e. $\tau (S)=1$) if and only if $A^+A=1$, namely if $A$ is isometric.
\end{proposition}
Finally Mackey's imprimitivity theorem allows one to find the most general form of a system of imprimitivity.
\begin{proposition}
Let $({\cal H'},S,{\cal G},\Lambda ,V,E)$ be a transitive system of imprimitivity. Let $q\in S$ be a generical element of $S$, for any $x\in S$ let $g_x\in \cal G$ be an element of $\cal G$ with the property $x=\Lambda (g _x) q$. Let $H _q$ be ``the little group", namely the closed subgroup of $\cal G$ defined by
\begin{equation}
g\in H _q\qquad \Leftrightarrow \qquad \Lambda (g)q=q.
\end{equation}
Then  one can represent $\cal H'$ as direct integral of Hilbert spaces on $S$ 
\begin{equation}
{\cal H'}=\int ^\oplus _S {\cal H'}(x)d\mu (x),
\end{equation}
where $d\mu (x)$ is a measure on ${\cal B}(S)$ having the same null sets as the spectral measure $E$. The vectors $\phi \in {\cal H'}$ can be represented by ``wave functions" $\psi (x)$ and  the projectors $E(I)$ as diagonal operators:
\begin{equation}
(E(I)\psi ) (x) = f_I(x)\psi (x),
\end{equation}
where $f _I(x)$ is the characteristic function of the Borel set $I \in {\cal B}$. 
Moreover the unitary representation $V$ of $\cal G$ takes the form of  an ``induced representation":
\begin{equation}
[V(g)\psi ](x)=\Bigg [{d\mu (x') \over d\mu(x)}\Bigg ]^{1\over 2}R(g_x^{-1}gg_{x'})\psi (x'),\qquad x'=\Lambda (g^{-1})x
\end{equation}
where $g_x^{-1}gg_{x'}\in H_q$ and $R$ is an unitary representation of $H _q$.
\end{proposition}

The introduction of a system of imprimitivity is very advantageous: 
in this way the probability that the result of the measurement, performed on the state $\phi$, belongs to the Borel set $I\in S$ takes the following simple form:
\begin{equation}
P(\phi , I)= \langle \phi, \tau (I)\phi \rangle=\langle A\phi, E(I)A\phi \rangle =\int _S f_I(x)\Vert \psi (x)\Vert ^2d\mu (x),\qquad \psi =A\phi .
\end{equation}
In other words the usual concept of  {\it probability density}, which can be found in the traditional formulation of quantum mechanics, can be re-established even if  spectral measures are replaced by generic POVMs.

The last step of our construction procedure is  the description of the most general intertwinig operator joining the imprimitivity system found by means of Mackey's theorem to the unknown covariance system. The following theorem allows one to know when such an operator exists and what its general form is. It is based on a generalization of an argument given in ref \cite{Dixmier2} and on Schur's lemma.
\begin{proposition}Let $\cal G$ be a locally compact topological group with a countable base of open sets and of type I. Let $\hat {\cal G}$ be its dual space, namely the space of equivalence classes of its irriducible representations.
Let $U$ and $V$ be two unitary representations of its, defined by their central decompositions:
\begin{equation}\label{primU}
U(g)=\int _{\widehat {\cal G}}^\oplus (U_\lambda (g)\otimes 1_\lambda )d\mu (\lambda ),\qquad  V(g)=\int _{\widehat {\cal G}}^\oplus (U_\lambda (g)\otimes 1'_\lambda )d\mu ' (\lambda ),
\end{equation}
acting respectively on Hilbert spaces
\begin{equation}\label{primH}
{\cal H}=\int _{\widehat {\cal G}}^\oplus {\cal H}_\lambda \otimes {\cal K}_\lambda  d\mu (\lambda ),\qquad {\cal H'}=\int _{\widehat {\cal G}}^\oplus {\cal H}_\lambda \otimes {\cal K'}_\lambda  d\mu ' (\lambda ),
\end{equation}
where $U_\lambda $ are irriducible representation and $1_\lambda$ and $1'_\lambda$ are the unity operators acting on the Hilbert spaces ${\cal K}_\lambda$ or ${\cal K'}_\lambda$.\\
An isometric intertwinig operator $A:{\cal H}\to {\cal H'}$
\begin{equation}
AU(g)=V(g)A,\quad \forall g\in {\cal G},\qquad A^+A=1
\end{equation}
exists if and only if $\mu $ is absolutely continuous with respect to $\mu'$ and 
\begin{equation}
 {\rm dim}({\cal K'}_\lambda )\geq {\rm dim}({\cal K}_\lambda )
\end{equation}
almost everywhere whith respect to $\mu$. In this case it will assume the following form:
\begin{equation} 
[A\phi ]_\lambda =\Bigg  ({d\mu\over d\mu'}\Bigg )^{1\over 2}(1_\lambda\otimes A_\lambda )\phi _\lambda ,\qquad A^+_\lambda A_\lambda =1,
\end{equation}
where $1_\lambda$ is the unity operator in ${\cal H}_\lambda$ and $A_{\lambda}:{\cal K}_\lambda\to {\cal K'}_\lambda $ is an isometry  defined almost everywhere with respect to $\mu$. \end{proposition}

The theorem reduces all our efforts, once we have the imprimitivity system, to the decomposition of V into I.U.R.s of the simmetry group $\cal G$.\\
The condition for the applicability of the theorem are not too restrictive, since most of the groups of physical interest have the required properties, namely they are locally compact with a countable base of open sets and of type I. However we shall see that the absolute continuity of the measure $\mu $ on $\hat{\cal G}$ with respect to $\mu '$ leads to interesting physical consequences, namely to a series of constraints on the realizability of some measurements on particular physical systems.


\section{Localization of events in space-time}
The first step necessary for  the description of a realistic quantum reference frame is the definition of its origin. From an operational point of view this is the description of the way in which a microscopical object can localize a point of the Minkowsky space-time manifold, namely an instant indicating the beginning of the time scale and a point in space with respect to which position measurements are referred. In other words, how a quantum system can point at a particular event, in a relativistic covariant way.\\
This kind of measurement can be described by a POVM on the Minkowsky space-time $\cal M$ covariant with respect to the universal covering of the proper orthochronous Poincar\'e group, which will be indicated by $\cal P$.
The problem has already been studied in \cite{Toller1} with a different method. We are going to rederive those results by means of the above developed construction procedure.\\ 
The first step is the construction of the most general imprimitivity system on $\cal M$ covariant with respect to $\cal P$, whose action $\tilde\Lambda$ on $\cal M$ is given by 
\begin{equation}
\tilde \Lambda (y,a)(x)=y+\Lambda (a)x,\qquad (y,a)\in {\cal P},\quad y\in {\cal T}_4,\quad a\in SL(2C),
\end{equation}
where $\Lambda :a\to \Lambda (a)$ is the representation of $SL(2C)$ acting on $\cal M$ by means of the Lorentz matrices.

The system of imprimitivity is transitive. If we choose as representative point of the only orbit in $\cal M$ under the action of $\cal P$ the origin $O=(0,0,0,0)$, we can recognize  the little group in the Lorentz group, or more precisely in $SL(2C)$, its universal covering. 
According to the imprimitivity theorem the unitary representation $V$ has the form of an induced representation
\begin{equation}
[V(y,a)\psi ](x)=D(a)\psi (x'),\qquad x,x' \in {\cal M},
\end{equation}
where $\psi $ takes its values in a Hilbert space $ \tilde{ \cal H}$, $D(a)$ is a unitary representation  (not necessarily irreducible) of $SL(2C)$ and 
\begin{equation}
x'=\Lambda (a^{-1})(x-y).
\end{equation}
The projection-valued measure $E$ on the homogeneous space $\cal M$ allows one to represent the vectors $\psi$ belonging to the Hilbert space $\cal H'$ as square-integrable vector-value function defined on $\cal M$.
The Lebesgue measure $d^4x$ on $\cal M$, canonically associated to Minkowsky coordinates, is invariant under the action of $\cal P$ and the norm of $\psi$ assumes the simple form:
\begin{equation}
\Vert \psi \Vert ^2=\int _{\cal M}\Vert \psi (x)\Vert ^2d^4x,
\end{equation}
while the spectral measure $E$ on the Borel $\sigma -$algebra {\cal B} of $\cal M$ assumes the diagonal form:
\begin{equation}
[E(I)\psi](x)=f_I(x)\psi (x),\qquad I\in {\cal B}.
\end{equation}

The second step is the decomposition of $V$ into I.U.R.s of the Poincar\'e group $\cal P$. 
We perform a Fourier transform on $\cal M$ and pass from the coordinate representation to the momentum representation: 
\begin{equation}
\tilde\psi (k)=(2\pi )^{-2}\int_{\cal M}\exp (ik\cdot x)\psi (x)d^4x,\qquad k\cdot x=x^\alpha k_\alpha,
\end{equation}
\begin{equation}
\Vert \psi \Vert ^2=\int \Vert \tilde \psi (k) \Vert ^2 d^4k.
\end{equation}
$V$ takes the following form:
\begin{equation}
[V(y,a)\tilde\psi ](k)=	exp (ik\cdot y)D(a)\tilde\psi (k'),\qquad k'=\Lambda (a^{-1})k.
\end{equation}
The physical states of $\cal H$ contain only non-negative energy representations, it follows that if $A$ is an intertwinig operator between $U$ and $V$, then $A{\cal H}\subseteq {\cal H''}\subseteq {\cal H'}$, where ${\cal H''}$ is the invariant subspace of ${\cal H'}$ which contains the vectors with non-negative energy, namely the wave functions $\tilde \psi (k)$ with support in the future cone $V _+$. We may disregard the values taken by $\tilde\psi (k)$ on the boundary of the cone, which has vanishing Lebesgue measure. In what follows we consider the subrepresentation $V''$ of $V$ acting on ${\cal H''}$.\\
Now we introduce for any $k$ in the open future cone an element $a_k\in SL(2C)$ defined by 
\begin{equation}\label{cono}
k=\Lambda (a_k)(M,0,0,0) ,\qquad {k^0}^2-{\bf k}^2=M^2,
\end{equation}
and the new wave function $\psi '$, defined by:
\begin{equation}
\tilde\psi (k)=D(a_k)\psi'(k).
\end{equation}
The representation $V''$ takes the following form:
\begin{equation}
[V''(y,a)\psi'](k)=\exp(ik\cdot y)D(a_k^{-1}aa_{k'})\psi '(k'),
\end{equation}
where $ a_k^{-1}aa_{k'}=u\in SU(2)$.
We can now consider the decomposition of $D$ into I.U.R.s of $SL(2C)$, whose matrix elements we indicate with $D^{\rho n}_{jmj'm'}(a)$. They are identified by two parameters: $\chi =(\rho ,n)$. Two different I.U.R.s identified by $(\rho ,n)$ and $=(\rho' ,n')$  are equivalent if and only if either $(\rho ,n) =(\rho' ,n')$, or $(\rho ,n) =(-\rho' ,-n')$. There are two series of I.U.R.s: the principal series with $\rho$ real and $n$ integer, and the supplementary series with $\rho$ imaginary and $n=0$ \cite{{GGV}, {Naimark}, {Ruhl}, {ST}}
. Moreover one should not forget the trivial one-dimensional representation. The restriction of these representations to the subgroup $SU(2)$ is given by 
\begin{equation}\label{repsl2c}
 D^\chi _{jmj'm'}(u)=\delta _{jj'}R^j_{mm'}(u),
\end{equation}
where $R_{mm'}^j(u)$ stands for the matrix elements of the I.U.R. of $SU(2)$, labelled by the integer or half-integer index $j$, with 
\begin{equation}
j=\vert{n\over 2}\vert, \vert{n\over 2}\vert +1,...\qquad m= -j, -j+1, ...,j-1,j 
\end{equation}
Every unitary representation of $SL(2C)$ can be decomposed uniquely into primary representations, which are direct sums of I.U.R.s, as $SL(2C)$ is a type I group. We consider the direct integral decomposition of the Hilbert space $\tilde{\cal H}$ into irreducible spaces labelled by the variable $\chi =(\rho,n)$, and introduce an index $\alpha $, which distinguishes the spaces where equivalent I.U.R.s operate:
\begin{equation}
\tilde {\cal H}= \int^\oplus _ {\widehat{SL(2C)} } \bigoplus _\alpha \tilde {\cal H}_\alpha^\chi d\omega (\chi ),
 \end{equation}
\begin{equation}
\Vert \psi \Vert ^2=\int _ {\widehat{SL(2C)} \times V _+}\sum _\alpha\Vert\psi _\alpha (k,\chi )\Vert ^2d\omega (\chi)d^4k,
\end{equation}
where $\omega$ is a generic measure on $\widehat{SL(2C)}$. \\
For fixed values of $\alpha, M, \chi $ the Poincar\'e group $\cal P$ acts in the way described by Wigner \cite{Wigner1}:
\begin{equation} 
[V''(y,a)\psi']_{\alpha ,jm}(k,\chi )=\exp (ik\cdot y)\sum _{m'}R^j_{mm'}(a_k^{-1}aa_{k'})\psi '_{\alpha ,jm'}(k',\chi ),
\end{equation}
as
\begin{equation} \label{repsu2}
a_k^{-1}aa_{k'}=u\in SU(2)
\end{equation}
Every I.U.R. of $\cal P$ with positive mass, identified by the variables (M,j), appears in the direct integral decomposition of $V''$ with a given multiplicity (defined almost everywhere on the positive real axis, i.e. on the $M$-axis). The multiplicity of a particular representation (M,j) is   strictly positive if the subset of $\widehat{SL(2C)}$, whose elements are the I.U.R.s $\chi =(\rho ,n) $ of $SL(2C)$ with $n\leq 2j$, has non-vanishing measure $\omega$. Then one can always assume that the multiplicity is as large as one needs, allowing the index $\alpha $ to take a sufficient number of different values. \par
As we have seen in the previous section, the intertwinig operator theorem can be applied once $U$, the unitary representation of $\cal P$ acting on the Hilbert space $\cal H$, is decomposed into direct integral of  spaces where I.U.R.s of $\cal P$ operate:
\begin{equation} \label{repres}
[U(y,a)\phi ]_{\alpha jm}(k)=\exp (ik\cdot y )\sum _{m'}R_{mm'}^j(u)\phi _{\alpha jm'}(k'),
\end{equation}
with 
\begin{equation}
\Vert\phi \Vert ^2=\int \sum _{\alpha ,j}\Vert\phi _{\alpha }(k,j)\Vert ^2 d\mu (k) .\end{equation}
The discrete index $\alpha $ distinguishes the spaces where equivalent I.U.R.s operate. Note that the range of the sum on the indices $\alpha$ and $j$ may depend on $M$. 
 The measure $d\mu (k) $ gives some informations about the mass spectrum of the system on which the measurement is performed. According to our third theorem, an isometric intertwinig operator $A$ between $U$ and $V''$ exists only if $\mu (k) $ (and therefore the corresponding measure on the range of $M$) is absolutely continuous with respect to the Lebesgue measure $d^4k$. This is possible if and only if the physical system on which the measurement is performed has a continuous mass spectrum, so we have to disregard the vacuum state  and the one-particle states, whose mass-spectrum has a vanishing Lebesgue measure. Moreover, if a value $j$ appears in the decomposition, the measure $\omega$ of the set $I\in {\cal B}(\widehat{SL(2C)})$ with $I= (\{\chi =(\rho ,n) \in \widehat{SL(2C)},\,  n\leq 2j\})$ has to be strictly positive.
Eventually, if these condition are satisfied the most general intertwinig operator takes the following form:
\begin{equation}
\psi '_{\alpha jm}(k,\chi )=\sum _{\alpha '}A^j_{\alpha \alpha '} (M,\chi ) \phi _{\alpha 'jm}(k),
\end{equation}
assuming that ${d\mu (k)\over d^4k}=1$ when $M$ belongs to the mass spectrum, with
\begin{equation}
\int \sum _\alpha  \overline {A^j_{\alpha \alpha '} (M,\chi ) }A^j_{\alpha \alpha ''} (M,\chi ) d\omega (\chi )=\delta _{\alpha '\alpha ''}.
\end{equation}
Finally the most general density of probability on the Minkowsky space-time, describing the measurement of the coordinates of an event individuated by a quantum state described by a vector $\phi \in {\cal H}$ takes the following form:
\begin{equation}\label{eventi1}
\rho (x)= \sum _\alpha \int _{\widehat{SL(2C)}} \Vert\psi _\alpha (x,\chi )\Vert ^2 d\omega (\chi ), 
\end{equation}
where
\begin{equation}\label{eventi2}
\psi _{\alpha pl}(x,\chi )= (2\pi )^{-2}\int \exp (-ik\cdot x)\sum _{\alpha ',jm}D^\chi _{pljm}(a_k) A^j_{\alpha \alpha '} (M,\chi ) \phi _{\alpha 'jm}(k)d^4k.
\end{equation}
This is the main result of ref \cite{Toller1}.

\section{Quantum frames of reference}
From an operational point of view a  reference frame $F$ can be defined by the operations which allow one to connect it to an initially  fixed frame $F_0$. In a relativistic theory the set of the allowed transformation is represented by the Poincar\'e group $\cal P$. Every element of $\cal P$ individuates the translation in the Minkowsky space-time and the Lorentz transformation which make the origin and the axes of the two frames coincide. We can also recognize in the four-vector individuating the translation the coordinates of the new origin with respect to the old one, while in the columns of the Lorentz matrix one finds the components of the new four orthogonal axes, relative to the old orthogonal basis. The ten indipendent parameters individuating the Poincar\'e transformation can also be recognized as relative observables of the two frames,  namely relative position, time, velocity and spatial orientation. Their description can be given by a POVM on $\cal P$,  covariant with respect to $\cal P$ itself. In other words the Poincar\'e group is in this case both  the symmetry group and the measure space, endowed with the invariant Haar measure $\nu$. The description  is simplified by the assumption of the classical nature of the frame $F_0$, in other words it will be considered an abstract mathematical tetrad with well-defined position, velocity and orientation. In this case the Poincar\'e group acts just on the system $F$ by means of left translation and the covariance condition assumes the following form:
\begin{equation}
U(g)\tau (I)U(g)^{-1}=\tau (gI),\qquad g\in {\cal P},\, I\subseteq {\cal P}.
\end{equation}
As shown in the previous  sections, the starting point is the application of Mackey's imprimitivity theorem. In this case it is particulary simple: there is only one orbit and, if we choose for example as representative point the identity $e\in {\cal P}$, the little group is reduced to the identity and the induced representation is simply the left regular representation or the direct sum of several representations equivalent to the left regular one and distinguished by the index $\alpha$:
\begin{equation} 
[V(g')\psi ]_\alpha (g)=\psi _\alpha (g'^{-1}g),
\end{equation}
\begin{equation} 
[V(y,a)\psi ]_\alpha (x,b)=\psi _\alpha (\Lambda (a^{-1})(x-y),a^{-1}b).
\end{equation}
It can be decomposed into i.u.r.s of $\cal P$ by means of the harmonic analysis on $\cal P$, defined by
\begin{equation}
\tilde \psi(\gamma  )=\int \psi(g)D^\gamma (g)d\nu (g),
\end{equation}
where $\gamma $ stands for  $(M,j)$. The inversion formula is given by:
\begin{equation}
\psi (g)=\int _{\widehat {\cal P}}{\rm Tr}[\tilde \psi(\gamma )D^\gamma (g^{-1})]d\hat\nu (\gamma),
\end{equation}
where $ d\hat\nu (\gamma)$ is the Plancherel measure on $\widehat {\cal P}$. 
On the new ``wave function ", defined on $\widehat {\cal P}$, the space of equivalence classes of i.u.r.s of $\cal P$, the group action assumes the following form:
\begin{equation} 
 [V(g)\tilde \psi ]_\alpha (\gamma )= D^\gamma (g)\tilde\psi _\alpha (\gamma ).
\end{equation}
This procedure can be repeated whenever the action of the symmetry group $\cal G$ on the measure space $S$ is free and transitive, namely if for all $ x,y\in S $ exists one and only one $ g\in {\cal G}$ so that $ y=\Lambda (g) x$. In this way, for example, we can construct time measurements covariant with respect to time translations and position measurements covariant with respect to space displacements.\\
In our case  we have just to combine the usual Fourier transform on $R^4$ and the harmonis analysis on $SL(2C)$, which give:
\begin{equation}
\tilde\psi _{\alpha ,jmj'm'}(k,\rho ,n )=(2\pi)^{-2}\int \exp (ikx)D^{\rho n}_{jmj'm'}(a)\psi _\alpha (x,a)d\mu (a)d^4x,
\end{equation}
and
\begin{equation}
\Vert \psi\Vert ^2=\int _{\widehat{\cal T}\times \widehat{SL(2C)}}\sum_\alpha {\rm Tr}[\psi _\alpha ^+(k,\chi)\psi _\alpha (k,\chi )]d^4kd\hat\mu (\chi),
\end{equation} 
where $d\hat\mu (\chi )$ is the Plancherel measure on $\widehat{SL(2C)}$, which is concentrated on the principal series only.\par
As in the preceding section we consider only $\cal H''$, the invariant subspace of $\cal H'$ which contains only the vectors with non-negative energy, and the subrepresentation $V''$ acting on $\cal H''$. For every $k$ belonging to the open future cone $V _+$ we introduce an element $a_k\in SL(2C)$ defined by equation (\ref{cono}) and the new wave function $\psi'$, given by:
\begin{equation}
\tilde\psi _{\alpha , plj'm'}(k,\chi )=\sum _{qs}D^\chi _{plqs}(a_k)\psi '_{\alpha ,qsj'm'}(k,\chi ).
\end{equation}
In this way the action of $\cal P$ on $\cal H''$ for fixed values of $\alpha, M, \chi, j',m'$  assumes the form introduced by Wigner \cite{Wigner1}:
\begin{equation} 
[V''(y,a)\psi']_{\alpha ,jmj'm'}(k,\chi )=\exp (ik\cdot y)\sum _{m'}R^j _{mn}(a_k^{-1}aa_{k'})\psi '_{\alpha ,jnj'm'}(k',\chi ),
\end{equation}
because of equation (\ref{repsu2}) and (\ref{repsl2c}) 
Once we have introduced the primary decomposition of $U$ shown in equation (\ref{repres}), the most general intertwinig operator can be found if and only if the measure $d\mu (k)$, defining the mass spectrum of the physical system on which the measurement is performed, is absolutely continuous with respect to $d^4k$. In this case, redefinig if necessary the normalization of the wave function $\phi \in {\cal H}$ so that ${d\mu (k)\over d^4 k }=1$ when $M$ belongs to the mass spectrum, we can write:
\begin{equation}
\psi '_{\alpha jmqs}(k,\chi )=\sum _{\alpha '}A^j_{qs,\alpha \alpha '} (M,\chi ) \phi _{\alpha 'jm}(k),
\end{equation}
with
\begin{equation}
\int \sum _\alpha \sum _{q=\vert {n\over 2}\vert}^\infty \sum _{s=-q}^q\overline {A^j(M,\chi )_{ \alpha \alpha ' qs }}A^j(M,\chi )_{\alpha \alpha '' qs }d\hat\mu (\chi )=\delta _{\alpha '\alpha ''}.
\end{equation}
Finally the density of probability assumes the following form:
\begin{equation}\label{frame1}
\rho (x,b)=\sum _\alpha \vert \psi _\alpha (x,b)\vert ^2,
\end{equation}
where
$$\psi _\alpha (x,b)  =(2\pi )^{-2}\int\exp (-ik\cdot x){\rm Tr}[D^\chi (b^{-1}a_k)\psi'_\alpha (k,\chi )]d^4kd\hat\mu (\chi) $$ $$=(2\pi )^{-6}\int d^4k \exp (-ik\cdot x)\int _0 ^{+\infty}d\rho\sum _{n=-\infty}^{+\infty}(n^2+\rho ^2)$$ 
\begin{equation}\label{frame2}
\times\sum _{j,q=\vert {n\over 2}\vert}^\infty \sum _{m=-j}^j\sum _{s=-q}^{q}D^{(\rho ,n)}_{qsjm}(b^{-1}a_k)\psi' _{\alpha jmqs }(k,\chi ).
\end{equation}
It is quite interesting to compare the density of probability on the Minkowsky space-time (\ref{eventi1}), which was found
indipendently in the previous  section, and the density of probability which can be found from (\ref{frame1}) by
integration on $SL(2C)$: 
\begin{equation}
\rho (x)=\int _{SL(2C)}\rho (x,a)d\mu (a),
\end{equation}
which after some calculations assumes the following form:
\begin{equation}
\rho (x)= \sum _\alpha \int {\rm Tr}[\psi' _\alpha (x,\chi )^+\psi' _\alpha (x,\chi )]d\hat\mu (\chi ), 
\end{equation}
with 
\begin{equation}
\hat\psi _\alpha (x,\chi )=(2\pi )^{-2}\int \exp (-ik\cdot x)\tilde \psi _\alpha (k,\chi )d^4k,
\end{equation}
where $d\hat\mu (\chi )$ is the Plancherel measure on the principal series of the i.u.r.s of $SL(2C)$.\\
As equations (\ref{eventi1}) and (\ref{eventi2}) show, the most general density of probability on the Minkowsky space-time admits a generic measure $d\omega (\chi) $ on the space of i.u.r.s of $SL(2C)$. From these considerations one can guess there are some constraints on the realizability of some measurement, whose properties can be found through an analysis of the physical meaning of the parametres $\chi =(\rho ,n)$ in this context. For example, as shown in ref \cite{Toller1}, one can recognize as ``baricentric" a measurement of events such that the  measure $\omega$ on ${\cal B}(\widehat{SL(2C)})$ appearing in equation (\ref{eventi1}) is concentrated on the trivial representation $D(a)=1$. Our results show that such a requirement can't be compatible with the measurement of the further parametres fully describing a reference frame. 
In other words the origin of the quantum reference frame can never be localized on the world line of the centre of mass of the microscopical system defining it. Moreover, as the same author suggested in a previous  paper \cite{Toller3}, for a complete description of a quantum reference frame kinematical variables are not sufficient, but internal degree of freedom must be involved. We can see for example that neither the invariant mass of the system nor its spatial distribution, namely the centre of mass position, can be arbitrarily fixed and disregarded, but have a foundamental role in the whole description.

\section{The observables relative to a quantum reference frame}
The quantum picture is complete if every classical element is disregarded and every kinematical variable of a quantum system $F_j$ is referred to a quantum reference frame $F_i$, namely a microscopical system with continuous mass spectrum. This can be simply obtained in two steps if the quantum systems $F_i$ and $F_j$ don't mutually interact.\\
First of all let's introduce as a preliminary tool a classical frame $F_0$, with respect to which
the parametres of the Poincar\'e tranformation connecting it to the quantum frame $F_i$ and a cinematical variable of the system $F_j$ are referred. The first ones are described by a POVM $\tau_i$ on the universal covering of the Poincar\'e group $\cal P$, acting on the Hilbert space ${\cal H}_i$, while the second ones are described by a POVM $\tau _j$ on a measure space $S$ acting on the Hilbert space ${\cal H}_j$. If there is no interaction the POVMs $\tau _i$ and $\tau _j$ and the unitary representation of $\cal P$ can be extended to the whole Hilbert space ${\cal H}={\cal H}_j\otimes {\cal H}_i$ by the relations:
\begin{equation}
U(g)=U_j(g)\otimes U_i(g),\qquad g\in {\cal P},
\end{equation}
\begin{equation}
\hat \tau _i (I)= I\otimes \tau _i(I) \qquad 
\hat \tau _j (J)= \tau _j(J) \otimes I,
\end{equation}
for all Borel subsets $I\subseteq {\cal P}$ and $J\subseteq S$.
One can easily see that $\hat \tau_i$ and $\hat \tau _j$  are endowed with the right covariance properties with respect to Poincar\'e transformations:
\begin{equation}\label{covfin1}
U(g)\hat \tau _i (I) U(g^{-1})=\hat \tau _i(gI)\qquad U(g)\hat \tau _j (J) U(g^{-1})=\hat \tau _j(\Lambda (g)J)\qquad \forall g\in {\cal P}.
\end{equation}
Moreover the operators in their ranges are mutually commuting:
\begin{equation}
[\hat\tau _i(I), \hat\tau _j(J)]=0,\qquad I\subseteq {\cal P}, J\subseteq S.
\end{equation}
If these condition are satisfied the {\it convolution} \cite{{Halmos},{Vak}} $\tau _{ij}$ of the two POVMs $\hat \tau _i$ and $\hat\tau _j$ can be defined by the relation:
\begin{equation}
\tau _{ij}(J)=\int f _J (\Lambda (g^{-1})x)d\hat \tau _i(g) d\hat\tau _j(x), \qquad g\in {\cal P},\quad x\in S,\quad J\subseteq S.
\end{equation}
It is suitable for the description of the relative observables of the system $F_j$ with respect to the quantum frame $F_i$. Indeed $\tau_{ij} $ is endowed with the properties of a POVM acting on the Hilbert space $\cal H$, namely positivity, $\sigma-$additivity and normalization. Moreover, as we expected, it is invariant under the action of the Poincar\'e group $\cal P$:
\begin{equation}
U(\tilde g)\tau _{ij} (J) U(\tilde g^{-1}) =\int f _J (\Lambda (g^{-1}\tilde g)\Lambda (\tilde g^{-1})x)d\hat \tau _i(g) d\hat\tau _j(x) =\tau _{ij} (J),
\end{equation}
in fact a Poincar\'e transformation will  act on both $F_i$ and $F_j$, changing the ``absolute'' cinematical variables of the two systems but leaving invariant the relative ones.\par
The mathematical description of the relations connecting two not-interacting quantum frames $F_i$ and $F_j$ can be obtained as a special case of this formalism. If in the previous discussion the quantum system $F_j$ has a continuous mass spectum, while the measure space $S$ coincides with the Poincar\'e  group again, $\tau _{ij}$ describes the measurement of the ten parametres of the transformation connecting the two frames. It assumes the following form:
\begin{equation}
\tau _{ij}(I)=\int f _I (g^{-1}g')d\hat \tau _i(g) d\hat\tau _j(g'),\qquad J\subseteq {\cal P}.
\end{equation}
The intuition can be helped by a calculation of the density of probability describing the statistics of the measurement, defined by 
\begin{equation}
\langle \phi_ i\otimes \phi_j,\tau _{ij}(I) \phi_ i\otimes \phi_j\rangle=\int _{\cal P}f_I(g)\rho _{ij}(g)d\mu (g),
\end{equation}
which assumes the following form:
\begin{equation}\label{prob}
\rho _{ij} (g)=\int _{\cal  P}\rho _i(\phi _i, g')\rho _j(\phi _j,g'g) d\mu (g'),
\end{equation}
where $\rho _i(\phi _i, g')$ and $\rho _j(\phi _j, g'g)$ are the densities of probability describing a measurement of the ``absolute'' parameters $g'$ and $g'g$, namely relative to a classical reference frame $F _0$, which were calculated in the previous section. In other words  if the Poincar\'e transformations identified by the elements $g'$ and $g'g$ connect the classical frame to the quantum frames $F_i$ and $F_j$ respectively, then the transformation from $F_i$ to $F_j$ will be individuated by the element $g\in {\cal P}$, whatever $g'$ may be. \par
The introduction of a third quantum frame $F_k$ in the description leads to some surprising consequences, which are commonly called ``the paradox of quantum frames''. While  the relative observables of $F_i$ and $F_j$, or of $F_i$ and $F_k$, can be described respectevely by $\tau _{ij}=\widetilde{\tau _i}*\tau _j$ or by $\tau _{ik}=\widetilde{\tau _i}*\tau _k$, the relative observables of $F _j$ and $F _k$ can't be obtained by the convolution $\widetilde{\tau _{ik}}*\tau _{ij}$ as it doesn't own the necessary properties. The operators in its range could be positive if and only if the POVMs $\tau _{ik}$ and $\tau _{ij}$ commute, but it can't be required and it is not generally true. One can easily see that the commutativity of the POVM $\{\tau _i (I)\}_{I\in \cal B (P)}$ is a sufficient condition for the commutativity of $\tau _{ik}$ and $\tau _{ij}$: 
\begin{equation}
[\hat\tau _i(I),\hat\tau _i(I')]=0 \quad\Rightarrow\quad [\tau _{ik}(I),\tau _{ij}(I')]=0,\qquad I,I'\subseteq {\cal P},
\end{equation}
however the first condition cannot be required. Note that the commutativity of the projectors in the range of the spectral measure $\{E _i(I)\}_{I\in \cal B(P)}$ does not involve the commutativity of the POVM $\tau _i(I)=A^+E_i(I)A$, unless the range of the intertwinig operator ${\cal H''}=A{\cal H}$ is an invariant subspace under the action of the projectors $E _i(I)$. One can easily see that in this case the positive operators in the range of the POVM $\tau _i$ would be projectors too, but this is forbidden by Pauli's theorem and by the non compatibility of the observables describing a reference frame. 
In other words sequential measurement of the relative parametres of two quantum frames $F_j$ and $F_k$ with respect to a third quantum frame $F_i$ are not compatible, even if they don't mutually interact \cite{{AK},{Toller3}}.

\section{Aknowledgments}
I am grateful to  M. Toller for his precious suggestions. I also wish to thank V. Moretti.

\end{document}